\documentclass[10pt,conference]{IEEEtran}
\IEEEoverridecommandlockouts
\usepackage{cite}
\usepackage{amsmath,amssymb,amsfonts}
\usepackage{algorithmic}
\usepackage{graphicx}
\usepackage{textcomp}
\usepackage{hyperref}
\usepackage{multirow}
\usepackage[table]{xcolor}
\usepackage[]{mdframed}
\usepackage{tikz}
\usepackage[caption=false, font=footnotesize]{subfig}
\usepackage{comment}
\usepackage{tcolorbox}
\def\BibTeX{{\rm B\kern-.05em{\sc i\kern-.025em b}\kern-.08em
    T\kern-.1667em\lower.7ex\hbox{E}\kern-.125emX}}

\mdfsetup{
    linewidth=.9pt
}

\newcommand{\rqone}{\textbf{RQ1.} What are the main categories of flaky failures?}
\newcommand{\rqtwo}{\textbf{RQ2.} Which failure categories are the most costly?}
\newcommand{\rqthree}{\textbf{RQ3.} How do the failure categories evolve over time?}
\newcommand{\rqfour}{\textbf{RQ4.} What are the priority flaky failure categories?}

\newlength\MAX  

\newcommand{\DrawPercentageBar}[1]{%
  \begin{tikzpicture}
    \fill[color=black!50]   (0.0 , 0.0) rectangle (#1*6ex , 1.5ex );
    \fill[color=black!15] (#1*6ex  , 0.0) rectangle (6.0ex, 1.5ex);
  \end{tikzpicture}%
}

\begin{document}

\title{

On the Diagnosis of Flaky Job Failures: Understanding and Prioritizing Failure Categories

}

\author{\IEEEauthorblockN{Henri A\"idasso}
\IEEEauthorblockA{\textit{Software and IT Engineering Department} \\
\textit{École de technologie supérieure}\\
Montreal, Canada \\
henri.aidasso.1@ens.etsmtl.ca}
\and
\IEEEauthorblockN{Francis Bordeleau}
\IEEEauthorblockA{\textit{Software and IT Engineering Department} \\
\textit{École de technologie supérieure}\\
Montreal, Canada \\
francis.bordeleau@etsmtl.ca}
\and
\IEEEauthorblockN{Ali Tizghadam}
\IEEEauthorblockA{
\textit{Network Softwarisation and AI} \\
\textit{TELUS}\\
Toronto, Canada \\
ali.tizghadam@telus.com}
}

\maketitle

\begin{abstract}
The continuous delivery of modern software requires the execution of many automated pipeline jobs. These jobs ensure the frequent release of new software versions while detecting code problems at an early stage. For TELUS, our industrial partner in the telecommunications field, reliable job execution is crucial to minimize wasted time and streamline Continuous Deployment (CD). In this context, flaky job failures are one of the main issues hindering CD. Prior studies proposed techniques based on machine learning to automate the detection of flaky jobs. While valuable, these solutions are insufficient to address the waste associated with the diagnosis of flaky failures, which remain largely unexplored due to the wide range of underlying causes. This study examines 4,511 flaky job failures at TELUS to identify the different categories of flaky failures that we prioritize based on Recency, Frequency, and Monetary (RFM) measures. We identified 46 flaky failure categories that we analyzed using clustering and RFM measures to determine 14 priority categories for future automated diagnosis and repair research. Our findings also provide valuable insights into the evolution and impact of these categories. The identification and prioritization of flaky failure categories using RFM analysis introduce a novel approach that can be used in other contexts. 
\end{abstract}


\section{Introduction}

Continuous Integration (CI) is a software engineering practice used to automate the build, testing, and deployment of code changes \cite{humble_continuous_2010} using a pipeline of jobs. Each job performs a specific task in the software delivery process, such as code compilation, unit testing, static code analysis, packaging, and deployment. The main benefits of CI pipeline jobs include faster delivery of code changes, more frequent releases, higher software quality, and reduced manual intervention \cite{hilton_usage_2016}. 
Successful job executions lead to immediate deployment of software updates, while failures signal code errors that developers must address by examining the job logs.

In principle, a job failure is assumed to reveal issues in the software code or the build script. In practice, however, pipeline jobs can fail because of unexpected reasons that are not necessarily related to development activities, causing the job to have a non-deterministic outcome on rerun, i.e. a \textit{flaky job failure}. These reasons can be of various kinds, including flaky tests \cite{bell_deflaker_2018, luo_empirical_2014}, CI environment issues, temporary network outages, and more \cite{zolfagharinia_not_2017, lampel_when_2021, olewicki_towards_2022}.

The high number of flaky failures at TELUS is causing mounting delays in production releases due to diagnosis and repair times. An analysis of jobs data collected from 80 projects shows that, on average, 1 in 4 failed jobs (25\%) is flaky. These flaky jobs mislead developers, who waste a considerable amount of time trying to understand the issue underneath, which is often out of the scope of their responsibilities. When uncertain of the result of a job, developers usually rerun it multiple times in the expectation of a change in outcome. 
Multiple reruns of jobs imply a significant waste of machine resources and developers' time, as they must wait for the reruns to complete \cite{olewicki_towards_2022}. At TELUS, diagnosing and repairing certain flaky failures requires the intervention of specialized teams, which causes additional delays. For instance, failures due to infrastructure problems such as CI servers can only be handled by the \textit{infrastructure team} that is responsible for managing these resources.

Prior studies mainly focused on approaches for automatically detecting flaky jobs to minimize the waste associated with multiple reruns. In particular, Lampel et al. \cite{lampel_when_2021} used jobs' telemetry data at Mozilla to train a Machine Learning (ML) model for detecting flaky job failures. At Ubisoft, Olewicki et al. \cite{olewicki_towards_2022} leveraged job log data to automate the detection of flaky jobs. Later, Moriconi et al. \cite{moriconi_automated_nodate} explored using Knowledge Graphs for the same purpose at Amadeus. Although multiple studies focused on identifying and predicting flaky test categories \cite{luo_empirical_2014, bell_deflaker_2018, rahman_flakesync_2024, fatima_flakyfix_2024} for automated diagnosis and repair, flaky tests are only a part of the issues. Despite the difficulties involved in diagnosing and repairing flaky job failures, to our knowledge, no prior study has focused on studying their root causes for automated solutions. 

There is a strong need to automate the diagnosis of flaky job failures to support developers in such a time-consuming activity. However, one of the biggest challenges is the wide variety of underlying causes of flaky jobs. Such diversity complicates understanding the root causes while requiring many specific repair solutions. As such, focusing on all categories would be too costly for organizations to implement, with limited benefits on categories of failure that rarely occur or occur relatively frequently but do not involve significant waste. In addition, many failure categories pose a challenge for automated diagnosis, as the limited number of examples in each category constrains the training of accurate ML models. 

A well-defined categorization and prioritization of flaky failures can help in training accurate ML models to detect specific categories of flaky failures as they occur. At a minimum, such a model will help developers to more quickly determine the category of failure they are facing and, therefore, the most appropriate team to solve the problem. Also, by prioritizing failure categories, one can focus efforts on the most recurrent and wasteful issues to implement targeted solutions for their repair (e.g., apply flaky test repair techniques \cite{bell_deflaker_2018, rahman_flakesync_2024} if the job failure falls under the flaky test category).

To address this gap, we analyze the different categories of flaky job failures at TELUS and then prioritize them based on Recency, Frequency, and Monetary measures. For this purpose, we identify and label 4,511 flaky job failures, which we use to answer the following research questions (RQs):

\textbf{\rqone} 40\% of the flaky job failures fall into the five most frequent categories, with \textit{misconfigured env variable} being the most prevalent, accounting for 15\% of occurrences.  

\textbf{\rqtwo} The five most costly categories include two among the five most frequent, namely \textit{misconfigured env variable} and \textit{job execution timeout}, and the most widespread across projects, which is \textit{container registry server error} affecting 60\% of the projects.

\textbf{\rqthree} Certain categories, such as the second most frequent \textit{docker daemon connection failure}, have not occurred for over a year. In contrast, others, like \textit{api gateway deployment error}, appeared only in the past year, and \textit{container registry server error} have become more regular over the past few months.

\textbf{\rqfour} We identified 14 priority categories for future efforts on automated diagnosis and repair, focusing on the most recent, frequent, and costly issues. We also outlined 14 other categories for our industrial partner to monitor as potential future threats.

Our main contributions are a taxonomy of flaky failure categories for diagnosis and an approach to prioritize these categories based on RFM measures. Also, the findings provide critical insights into the impact and dynamics of different flaky failure categories at TELUS while laying the foundation for future work on automated diagnosis and repair of flaky jobs. A replication package \cite{aidasso_diagnosis_2025} will be made publicly available.

\section{Background and Motivation}

\subsection{CI/CD at TELUS}

TELUS is a leading Canadian telecommunications company that offers a broad range of products and services with applications across various sectors, including networking, healthcare, agriculture, and more. TELUS develops a wealth of software applications, such as Software-Defined Networks (SDNs), which manage dynamic traffic and ensure security in large-scale networks. The development of such software involves several specialized teams such as the \textit{network team} responsible for hardware and network configurations (e.g., virtual private networks \cite{ezra_secured_2022}); the \textit{development team}, who handle software development and maintenance; and the \textit{infrastructure team}, who focus on providing and maintaining infrastructure resources for deployments and production management.

At TELUS, the continuous deployment of code changes relies on a variety of distributed tools, which creates an environment more susceptible to flaky failures. Developers submit their changes to a self-hosted instance of the GitLab platform, which is also responsible for scheduling the execution of build jobs. In general, the build jobs automate tasks such as compilation, unit/integration testing, container image creation, image security scanning, and deployment. To run the build jobs on GitLab, the infrastructure team set up several self-managed GitLab Runners\footnote{\url{https://docs.gitlab.com/runner/}} in Kubernetes pods\footnote{Kubernetes pods are computing resources composed of one or more containers, capable of running applications or specific tasks within an isolated Linux environment \cite{noauthor_pods_nodate}.}. These pods are created on a self-hosted Container Platform (CP) used to manage infrastructure resources. During execution, the jobs call on various services, such as the self-hosted web-based static analysis tool for code quality checks, the internal container registry for pulling and pushing container images, public cloud platforms for deployments, and more. In addition, most jobs leverage container tools such as Docker to ensure job execution in an isolated environment or build a new software image version. When a job fails, developers analyze the execution logs to diagnose the issue. If they cannot understand or resolve the problem based on the logs, the developers submit a ticket to the team they believe is best suited to handle the issue (e.g., the infrastructure team).

\subsection{Flaky Job Failures}

\begin{figure}
  \begin{center}
      \includegraphics[width=.7\linewidth]{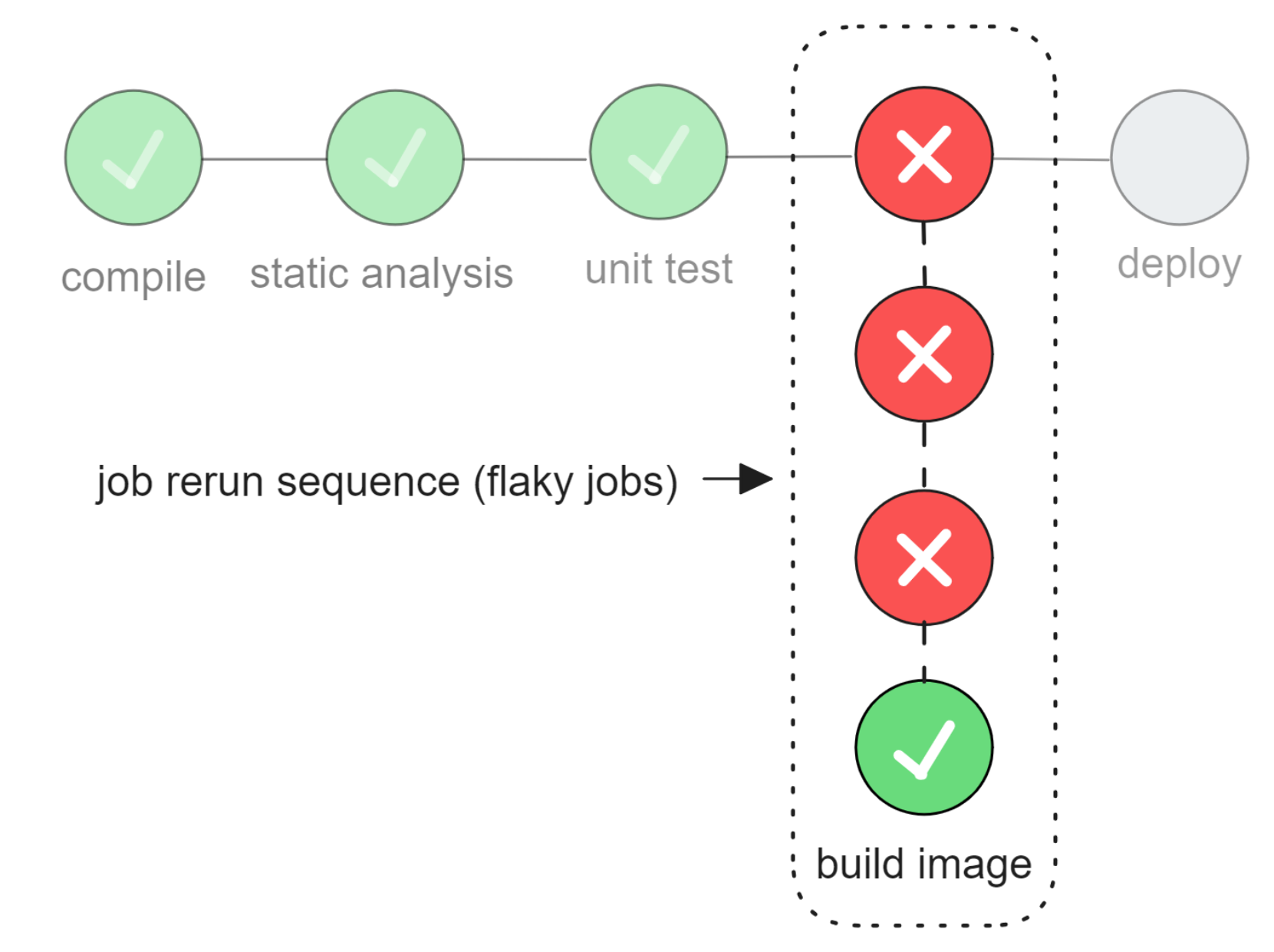}
  \end{center}
\caption{Example of a pipeline, highlighting the rerun sequence of the \textit{build image} job. After an initial failure, the \textit{build image} job is rerun three times before passing, revealing the flakiness of the job failures in the rerun sequence.}
\label{fig:rerun_suite}
\end{figure}

A flaky job failure is a non-deterministic failure caused by irregular issues, typically related to the CI environment, rather than coding errors \cite{olewicki_towards_2022, durieux_empirical_2020}. Such failures intermittently pass or fail upon multiple reruns without any source code or build script changes. As a result, they hinder the reliability of pipeline jobs and can be very difficult to detect and diagnose \cite{lampel_when_2021, olewicki_towards_2022, moriconi_automated_nodate}. 
Figure~\ref{fig:rerun_suite} illustrates the exhibition of flaky job failures in a job rerun sequence, including the initial job failure.

The frequent occurrence of flaky failures leads to a substantial waste of time and resources invested in their diagnosis and repair. A preliminary analysis of job data collected across 80 TELUS projects reveals that one in four job failures (25\%) is identified as flaky upon reruns. In the best cases, flaky failures can be \textit{repaired}, i.e. turned into success, simply by rerunning the job several times without further diagnosis. For instance, a job that fails due to an overloaded CI server can pass if rerun later when the server load has decreased. So, in doubt, developers systematically rerun any job failure with unclear results or that they suspect to be flaky. Such a practice results in inefficient use of machine resources and adds a heavy load to already overloaded CI servers \cite{olewicki_towards_2022}. 

In more complicated cases, diagnosing and repairing flaky failures requires the intervention of specialized teams, which can extend over several days. For example, a job failure that appears to be caused by a domain name resolution issue may, in reality, be masking an underlying infrastructure problem. Hence, developers submit a ticket to the team they deem the most appropriate between the \textit{network team} and the \textit{infrastructure team} for diagnosis. As tickets are handled in sequence, this request can take a long time (up to 48 hours, according to interviews) before being processed. In the worst cases, solving the problem requires multiple trial-and-error attempts. The time spent on these failures is even greater if the developers initially direct the ticket to the wrong team.

\subsection{Why State-of-the-Art Does Not Suffice}

The organization of the CI/CD system at TELUS and interviews with engineers have enabled us to identify three important points that motivate our present study:

\begin{itemize}
    \item TELUS does not have dedicated teams, such as Mozilla's Sheriffs \cite{lampel_when_2021}, to diagnose flaky job failures upon detection. So, while automated detection approaches proposed in previous studies \cite{lampel_when_2021, olewicki_towards_2022, moriconi_automated_nodate} are valuable, they remain insufficient to address these failures at TELUS. So, there is a strong need for automated diagnostic solutions to mitigate the associated time and resource inefficiencies.

    \item TELUS uses a wide range of distributed tools for its pipeline jobs, which, combined with the development of network applications, makes its CI system more susceptible to specific flaky failures, such as different types of networking issues. %

    \item  The large variety of flaky job failure causes (extending far beyond flaky tests) pose a challenge for automated diagnosis and repair, as addressing all of them is impractical. Some failures are rare, making it challenging to train accurate classification models without enough examples, while others require extensive investigation, often spanning several days at TELUS.
\end{itemize}

For these reasons, we aim to identify and prioritize the various categories of flaky failures. At a minimum, our findings will assist developers in diagnosing flaky failures more rapidly (e.g., direct a ticket to the right team). At best, the prioritization of flaky failure categories will enable us to build, in future research, a classification model to automate diagnosis and trigger specific repair actions for each priority category. 

\section{Methodology}

\subsection{Research Questions}
The main goal of this study is to prioritize the different categories of flaky failures to facilitate their automated diagnosis and repair. For this purpose, we adopted an approach based on RFM Analysis \cite{funatsu_data_2011}, a data mining method commonly used in marketing to segment and prioritize customers based on Recency (R), Frequency (F), and Monetary (M) measures. Hence, we first identify the different categories of flaky failures and study their frequency (RQ1). Then, we evaluate the monetary cost associated with these categories (RQ2) and analyze how each evolves over time (RQ3). Next, we build an RFM model to segment the identified categories based on their recency, frequency, and monetary cost measures (RQ4). This segmentation is analyzed to identify categories that should be prioritized for automated diagnosis and repair in future research. To achieve this, we design the following research questions (RQs):

\begin{itemize}
    \item[] \rqone
    \item[] \rqtwo
    \item[] \rqthree
    \item[] \rqfour
\end{itemize}

\begin{table}
\caption{Overview of the Collected Data}
\begin{center}
\begin{tabular}{l r @{}}
\hline
\textbf{Metric} & \textbf{Value} \\
\hline
    Start Date & February 1, 2018 \\
    End Date & July 11, 2024 \\ 
\hline
    \# Projects & 80 \\
    \# Languages & 13 \\
    \# Commits & 45,545 \\
    \# Jobs & 157,496 \\ \hline
    \# Success Jobs & 109,509 (69.53\%) \DrawPercentageBar{0.695} \\ 
    \# Failed Jobs & 31,058 (19.72\%) \DrawPercentageBar{0.197} \\
    \# Skipped Jobs & 8,443 (5.36\%) \DrawPercentageBar{0.054} \\
    \# Canceled Jobs & 5,804 (3.69\%) \DrawPercentageBar{0.037} \\
    \# Manual Jobs & 2,588 (1.64\%) \DrawPercentageBar{0.016} \\
    \# Created Jobs & 94 (0.06\%) \DrawPercentageBar{0.006} \\ 
    \hline
    \# Rerun Jobs & 25,241 (17.96\%) \DrawPercentageBar{0.179} \\ 
    \# Flaky Jobs & 16,378 (11.65\%) \DrawPercentageBar{0.116} \\ 
    \# Flaky Job Failures & 7,763  (25\%) \DrawPercentageBar{0.249} \\
\hline
\end{tabular}
\label{tab:collected_data}
\end{center}
\end{table}

\subsection{Data Collection}

Table~\ref{tab:collected_data} presents a summary of the collected data. In collaboration with our industrial partner, we selected 14 GitLab groups where developers actively contribute to major TELUS projects. After filtering out inactive and toy (with less than 50 commits) projects, we identified 80 projects varying in size, maturity, and purpose (e.g., network applications, data analytics, etc.) and covering 13 programming languages. We extracted the entire build job history of the 80 projects over 6.5 years, from February 1st, 2018, to July 11th, 2024, using the GitLab REST APIs\footnote{https://docs.gitlab.com/ee/api/jobs.html}. As a result, we collected 157,496 jobs' JSON metadata and 31,058 failed job logs (1.3GB). Then, we consolidated the JSON metadata into a CSV file and filtered out non-completed jobs (e.g., \textit{canceled}, \textit{skipped}) as they are not relevant to this study. We obtained 140,567 \textit{success} and \textit{failed} jobs.

We identified flaky jobs based on the following heuristic: if a job is rerun with no changes to the source code (i.e., on the same commit) and its outcome changes between success and failure, it is considered as flaky; otherwise, it is treated as a regular job. This approach was used previously \cite{olewicki_towards_2022} and aligns with the development practices at TELUS, as confirmed through interviews. Hence, we grouped the jobs by \textit{project}, \textit{name}, and \textit{commit} and identified the job rerun sequences (i.e., the groups of jobs with at least two executions of the job on the same commit). We saved these rerun sequences data for further analysis. For each job rerun sequence containing at least one \textit{success} and one \textit{failed} job, we labeled its jobs as flaky. As a result, we found 16,378 flaky jobs from which we excluded the jobs with a \textit{success} status. 

\textbf{We obtained 7,763 flaky job failures representing a quarter (25\%) of the failed jobs} as shown in Table~\ref{tab:collected_data}. Out of these 7,763 flaky job failures, we found that 2,507 (33.29\%) jobs have missing log data. A brief investigation revealed that this phenomenon stems from version compatibility issues between self-hosted GitLab and GitLab Runners following major upgrades, which result in job execution logs not being recorded \cite{4ces_answer_2021}. Engineers at TELUS confirmed that the lost logs resulted from a major migration of GitLab to a public cloud in early 2022. Since our approach to identifying the different categories of flaky failures is mainly based on log data, we excluded the flaky failures with missing logs.

\textbf{The 5,256 flaky job failures with log data constitute the initial dataset for the current study}. We label this dataset of flaky failures using the approach described below. In addition, we use the rerun sequences data to answer RQ2. For confidentiality reasons, our replication package does not include the datasets.

\begin{figure*}[tb]
  \begin{center}
      \includegraphics[width=.9\textwidth]{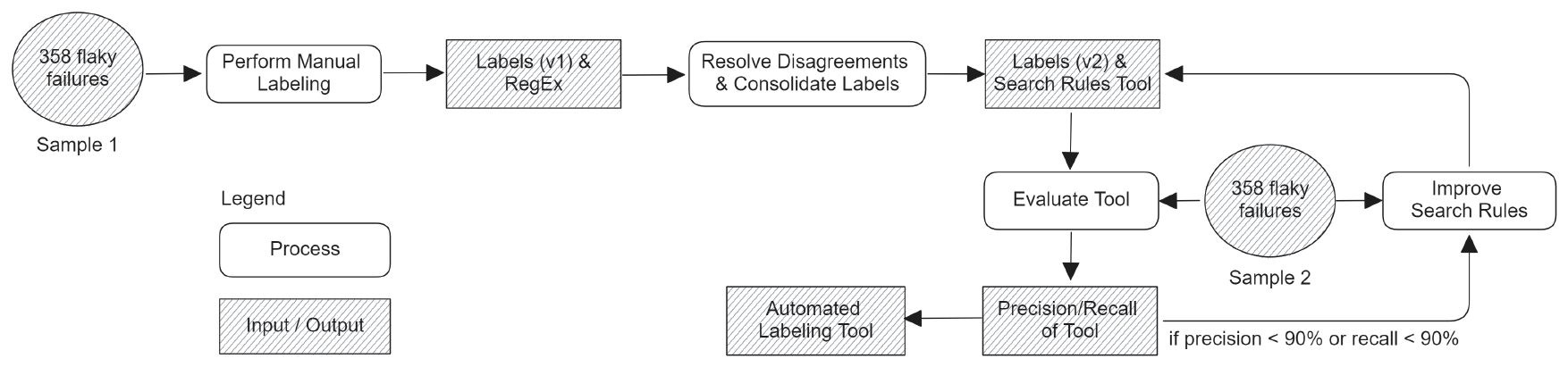}
  \end{center}
\caption{Overview of the approach for building an automated flaky failure labeling tool.}
\label{fig:labeling_tool}
\end{figure*} 

\subsection{Approach for Identifying Flaky Failure Categories}
\label{sec:categorization_approach}

To identify the different categories of flaky failures (i.e., data labeling), we use a semi-automated approach involving (1) the development of a labeling tool based on search rules and (2) the application of this tool to label the dataset systematically. A labeling tool is essential for this study to complete the labeling of thousands of flaky job failures. It can also be used in future studies to obtain labeled data. Figure~\ref{fig:labeling_tool} presents our approach to developing the automated labeling tool.

\textbf{Dataset Sampling.} Manually analyzing thousands of job logs for labeling is not practical. So, similarly to previous studies \cite{beller_modern_2014, ghaleb_studying_2019}, we use two statistically significant samples of 358 job logs each, out of the 5,256 in the initial dataset (a confidence level of 95\% and a 5\% margin of error). Sample 1 is used for manual labeling and building the automated labeling tool, while Sample 2 is reserved to evaluate the tool's accuracy. We randomly select from the initial dataset 358 flaky failures constituting Sample 1, while Sample 2 is drawn from the dataset deprived of Sample 1 jobs to ensure that the tool is validated on unseen data.

\textbf{Sample Manual Labeling.} For each job of Sample 1 that is not yet labeled, we manually analyze the logs to identify the cause of the failure. When the error message is initially unclear, we use StackOverflow (SO) to understand it better. For example, an investigation on SO has enabled us to realize that the \texttt{"exit code 137"} error indicates that a container ran out of memory due to the limits specified in its pod \cite{marshall_answer_2021}. Then, we use open coding \cite{khandkar_open_2009} to assign the flaky failure to an existing category or create a new one with a name reasonably indicative of the failure root cause (e.g., \textit{container oom error} for the previous error example). Each category name has been validated with TELUS engineers to ensure clarity. If a new category is created, we look in the logs for a regular expression (regex) capable of identifying failures with highly similar logs as precisely as possible to maximize the precision of our tool. We then use a script that searches for similar job failures based on the logs using this new regex. The jobs found are then also assigned to the newly created category. To avoid mislabeling, we manually validate each new job labeled with the new category name. Also, the regex pattern and the associated label are saved for later building our search rules-based tool. We repeat this process until all the jobs in Sample 1 are labeled. As a result, we obtain the first version of labels (v1) and their associated regexes.

\textbf{Labeling Tool Building.} To develop the labeling tool, we first established the final set of labels (v2) through regular meetings with TELUS engineers, during which we discussed and resolved disagreements. Additionally, we consolidated similar homogeneous categories to avoid excessive labels with limited examples, which would hinder efficient data analysis and the training of future ML classifiers. For instance, we merged the \textit{missing env variable} category with the  \textit{misconfigured env variable} category since they have similar log error messages, and the diagnosis and repair of both these issues consist of finding and setting the correct environment variable value. Each time we merge categories, we combine the corresponding regexes (e.g., using the or $\vert$ operator). In the end, we defined a set of 51 heuristics in the form of search rules that associate a regex pattern to one of the 46 final labels. We then developed a Python script to leverage these search rules to automatically label an input dataset of flaky failures with the appropriate labels using the job logs.

\begin{table}
\caption{Labeling Tool Evaluation Results}
\begin{center}
\begin{tabular}{llll}
\hline
\textbf{Input} & \textbf{Iteration} & \textbf{Recall} & \textbf{Precision} \\
\hline
    Sample 1 & - & 99.44\% & 100.0\% \\ \hline
    Sample 2 & 0 & 81.56\% & 99.31\% \\ \hline
    Sample 2 & 1 & 91.34\% & 99.31\% \\
\hline
\end{tabular}
\label{tab:evaluation_results}
\end{center}
\end{table}

\textbf{Tool Evaluation and Improvement.} We evaluate the performance of our labeling tool using two metrics: recall and precision. Recall is defined as $\frac{n_{labeled}}{N}$, where $N$ is the total size of the input dataset and $n_{labeled}$ is the number of jobs labeled by the tool. Precision is calculated as $\frac{n_{correct}}{n_{labeled}}$, where $n_{correct}$ represents the number of jobs correctly labeled, determined through manual inspection. To ensure that our labeling tool works correctly on familiar data, we first evaluated its performance on Sample 1. The results obtained, as reported in Table~\ref{tab:evaluation_results}, show that the tool is as effective as manual labeling, with a precision of 100\% and a recall of 99.44\%. There were two jobs for which we could not determine the cause, so we manually labeled them as \textit{unknown failure}. These were the only jobs the tool was also unable to label.

On the unseen data from Sample 2, the labeling tool achieved a precision of 99.31\%, demonstrating our approach's effectiveness in accurately labeling flaky job failures as intended. Also, the obtained recall of 81.56\% on Sample 2 shows that the identified categories are pretty representative of the types of failures in the overall dataset. Although we obtained a satisfying recall, we sought to improve it using the remaining unlabeled data from Sample 2. So, we followed the manual labeling process again, updating the regex patterns and search rules after each iteration. In the end, no new categories were discovered. We stopped after achieving precision and recall scores of over 90\%. As a result, we increased the tool's recall to 91.34\% on Sample 2.

\textbf{Automated Labeling.} We applied the labeling tool to our 5,256 flaky job failures dataset, successfully labeling 85.83\% of the dataset, i.e. 4,511 jobs. We consider these labeled flaky failures to be representative of the issues plaguing CI at TELUS. Since our study focuses on diagnosing flaky failures, we limited our analysis to failures where the root cause can be determined from the logs and successfully labeled using the tool. Consequently, the remaining 745 unlabeled failures have been excluded from the analysis.

\begin{mdframed}
\textit{Our labeling tool achieves a precision score of 91\% on an unseen representative sample. Using this tool, \textbf{we labeled 4,511 flaky failures} (86\% of the initial dataset). We then use the labeled dataset to answer our RQs.}
\end{mdframed}

\section{Results}

\subsection{\textbf{\rqone}}

\textbf{Motivation}. This RQ aims to identify the most prevalent categories of flaky failures. Despite efforts invested in research on flaky jobs, there is a lack of awareness about specific categories of flaky failures that can be used for diagnostic purposes. Also, the findings of this RQ will provide valuable insights into the recurrent categories of flaky failures that we need to consider prioritizing to maximize the impact of automated diagnosis and future repair solutions.

\textbf{Approach.} To answer this RQ, we analyze our dataset of 4,511 flaky failures, labeled using the approach described in Section~\ref{sec:categorization_approach}. We report statistics about the number and proportion of flaky job failures in each identified category. In addition, we report the number of distinct projects affected by each failure category to better assess their impact across projects. Also, to facilitate the understanding of the flaky failure categories, we use the card sorting \cite{spencer_card_2009} technique to organize related failure categories into meaningful groups.

\begin{table}[]
\caption{Categories of Flaky Job Failures at TELUS}
\begin{scriptsize}
\begin{tabular}{p{2.3cm}p{3.4cm}p{.4cm}p{.4cm}p{.2cm}}
\hline
\textbf{Group}                              & \textbf{Category}                   & \textbf{\#} & \textbf{\%} & \textbf{P$^{\mathrm{a}}$} \\ \hline

Environment Variables                       & misconfigured\_env\_variable        & \cellcolor{black!15}{673}         & 14.92       & \cellcolor{black!15}{35}               \\ 
                                            \hline
\multirow{7}{*}{Container Issues}           & container\_already\_exists          & 2           & 0.04        & 1                \\
                                            & container\_not\_found               & 31          & 0.69        & 7                \\
                                            & docker\_daemon\_connection\_failure & \cellcolor{black!15}{325}         & 7.2         & 13               \\
                                            & image\_build\_permission\_denied    & 8           & 0.18        & 1                \\
                                            & image\_build\_read\_error           & 17          & 0.38        & 5                \\
                                            & image\_push\_write\_error           & 1           & 0.02        & 1                \\
                                            & image\_security\_scan\_failure      & 8           & 0.18        & 7                \\ 
                                            \hline
\multirow{4}{\linewidth}{Unauthorized Access}        & certificate\_verification\_failure  & 56          & 1.24        & 16               \\
                                            & container\_platform\_auth\_failure                  & \cellcolor{black!15}{314}         & 6.96        & \cellcolor{black!15}{34}               \\
                                            & repository\_access\_denied          & 29          & 0.64        & 9                \\ 
                                            \hline
\multirow{6}{*}{Limits Exceeded}            & docker\_pull\_limit\_reached        & 30          & 0.67        & 5                \\
                                            & cloud\_token\_limit\_exceeded      & 39          & 0.86        & 6                \\
                                            & job\_execution\_timeout             & \cellcolor{black!15}{306}         & 6.78        & 19               \\
                                            & remote\_call\_timeout               & 76          & 1.68        & 22               \\
                                            & runner\_pod\_waiting\_timeout       & 199         & 4.41        & \cellcolor{black!15}{34}               \\
                                            & stuck\_or\_timeout\_failure         & 14          & 0.31        & 9                \\ 
                                            \hline
\multirow{6}{\linewidth}{Remote Resource Issues}     & api\_gateway\_deployment\_error           & 161        & 3.57        & 6                \\
                                            & image\_not\_found        & \cellcolor{black!15}{221}         & 4.9         & 25               \\
                                            & container\_registry\_server\_error  & 213         & 4.72        & \cellcolor{black!15}{50}               \\
                                            & external\_file\_invalid\_format     & 125         & 2.77        & 8                \\
                                            & http\_resource\_not\_found          & 38          & 0.84        & 6                \\
                                            & service\_unavailable                & 2           & 0.04        & 2                \\ 
                                            \hline
\multirow{5}{\linewidth}{Networking Issues}          & connection\_closed\_reset\_broken   & 204         & 4.52        & \cellcolor{black!15}{38}               \\
                                            & connection\_refused                 & 44          & 0.98        & 11               \\
                                            & host\_resolution\_failure           & 91          & 2.02        & 29               \\
                                            & broker\_connection\_failure         & 7           & 0.16        & 2                \\
                                            & ssl\_connection\_error              & 16          & 0.35        & 5                \\ 
                                            \hline
Flaky Tests                                 & flaky\_test                         & 179         & 3.97        & 11               \\ 
                                            \hline
\multirow{5}{\linewidth}{Infrastructure Issues}      & helm\_resource\_error               & 99          & 2.19      & 33               \\
                                            & runner\_image\_pull\_failure        & 99          & 2.19        & 28               \\
                                            & runner\_instance\_error             & 32          & 0.71        & 8                \\
                                            & runner\_pod\_failure                & 154         & 3.41        & 32               \\
                                            & runner\_pod\_not\_found             & 10          & 0.22        & 7                \\ 
                                            \hline
\multirow{4}{\linewidth}{Dependency Issues}          & buggy\_dependency                   & 13          & 0.29        & 4                \\
                                            & dependencies\_conflict\_error       & 129         & 2.86        & 4                \\
                                            & dependency\_installation\_failure   & 124         & 2.75        & 22               \\
                                            \hline
Transient VCS Errors                       & git\_transient\_error               & 113         & 2.5         & 31               \\ 
                                            \hline
\multirow{3}{*}{Memory Issues}              & container\_oom\_error               & 55          & 1.22        & 11               \\
                                            & testing\_device\_oom\_error         & 6           & 0.13        & 5                \\
                                            & static\_analysis\_tool\_oom\_error               & 57          & 1.26        & 8                \\ 
                                            \hline
\multirow{2}{\linewidth}{Repository File Issues} & repository\_file\_access\_error     & 60          & 1.33        & 10              \\
                                            & repository\_file\_not\_found        & 37          & 0.82        & 10               \\ 
                                            \hline
\multirow{2}{\linewidth}{Internal OS Issues}         & apt\_timezone\_issue                & 66          & 1.46        & 7               \\
                                            & os\_cmd\_execution\_error           & 12           & 0.27        & 3                \\ 
                                            \hline
Database Issues                             & db\_table\_undefined                & 16          & 0.35        & 1             \\

\rowcolor{black!10}\multicolumn{2}{l}{\textbf{Total}} & \textbf{4,511} & \textbf{100\%} & \textbf{80} \\
                                            \hline
                                    
\multicolumn{5}{l}{$^{\mathrm{a}}$Number of projects affected.}
\end{tabular}
\label{tab:categories}
\end{scriptsize}
\end{table}

\textbf{Results.} Table~\ref{tab:categories} presents the 46 identified categories of flaky failures organized into 14 thematic groups. For each category, it shows the frequency, the percentage proportion, and the number of projects impacted. Additionally, it highlights in gray the top five categories with the highest frequencies and the top five with the highest number of affected projects (i.e., the most popular across projects).

\textbf{The most frequent category is related to misconfiguration of environment variables, accounting for $\approx$ 15\% of the flaky failures.} Our analysis reveals that of the 4,511 labeled flaky job failures, 673 (14.92\%) belong to the \textit{misconfigured env variable} category. This failure category encompasses missing environment variables, misspelled variable names, and incorrect values of environment variables used during job execution. The \textit{misconfigured env variable} category is also the third most widespread category across the projects, affecting 35 projects of the 80 studied. This result underscores the critical role of managing CI environment variables in development activities. However, CI variable changes are not version-controlled, thus not accounted for in prior flakiness studies, calling into question the commonly accepted assumptions about flaky jobs (e.g., no changes made by developers).

\textbf{Over 40\% of the flaky job failures belong to the top five most frequent failure categories.} Right below the most frequent category \textit{misconfigured env variable}, the next categories are \textit{docker daemon connection failure}, \textit{container platform auth failure}, \textit{job execution timeout}, and \textit{image not found}, which account respectively for 7.2\%, 6.96\%, 6.78\%, and 4.9\% of the studied flaky failures. Collectively, these five most frequent categories constitute 40.76\% of the flaky job failures, highlighting the need for prioritization.

\textbf{The most popular category across projects is the {container registry server error} affecting over 60\% of the projects and ranking 6th most frequent}. Our results show that the \textit{container registry server error} is the most widespread failure category across projects, appearing in 50 of the 80 studied (62.5\%). 
It ranks just below the top five most frequent categories, accounting for 4.72\% of the flaky failures. This failure category occurs when a server-side issue prevents jobs from successfully pulling (or pushing) images from the shared container registry. 

Furthermore, the most popular categories across projects also rank highly in frequency. For instance, two of the top five most popular categories across projects, \textit{misconfigured env variable} and \textit{container platform auth failure}, are among the top five most frequent. The remaining three most popular categories highlighted—\textit{container registry server error}, \textit{connection closed reset broken}, and \textit{runner pod waiting timeout}—are the 6th, 7th, and 8th most frequent, representing 4.72\%, 4.52\%, and 4.41\% of flaky failures, respectively.

\begin{mdframed}
    \textit{\textbf{Answer to RQ1.} The five most frequent categories cover more than 40\% of the flaky job failures and are  \textbf{misconfigured env variable} (14.92\%), \textbf{docker daemon connection failure} (7.2\%), \textbf{container platform auth failure} (6.96\%), \textbf{job execution timeout} (6.78\%), and \textbf{image not found} (4.9\%). The 6th, \textbf{container registry server error} (4.72\%), affects the most number (60\%) of projects.}
\end{mdframed}

\subsection{\textbf{\rqtwo}}

\textbf{Motivation.} The objective of this RQ is to determine which failure categories are the most costly in terms of diagnosis time and machine resources due to reruns. In particular, we sought to determine whether the most frequent categories are the most costly. The results of this RQ will provide our industrial partner with valuable insights into the most wasteful failure categories, enabling a better prioritization for automated diagnosis and repair solutions.

\textbf{Approach.} As discussed before, the waste associated with flaky failures is related to the multiple job reruns and the (long) time spent by engineers diagnosing the initial failure. Therefore, to evaluate the cost associated with each failure category, we design a cost model including two components: (1) the machine cost and (2) the diagnosis cost. It is inspired by an existing approach for estimating the cost of executing and diagnosing test failures at Microsoft \cite{herzig_art_2015}. 

To estimate the total machine cost ($Cost_{machine}$) associated with a failure category, we calculate the cumulative duration spent by the jobs in the category. Then, we apply the average pricing rate of the infrastructure machines used to execute the jobs. Although this pricing rate can vary based on factors such as region and instance type, for simplicity, we assume all jobs run on the equivalent of a high-memory virtual machine (\textit{n2-highmem-128}) from Google Cloud. We calculated the pricing rate to be \$0.14/min. Finally, the formula for calculating the machine cost is outlined in Eq~\ref{eq:cost_machine} as follows:

Let $E=\{job_{1},job_{2},...,job_{n}\}$ denote the ensemble of $n$ flaky job failures within the failure category $E$, each with respective durations $d_{1},d_{2},...,d_{n}$ in minutes (i.e., $d_{i} = job_{i}[\mathrm{duration}], i=1..n$).

\begin{equation}
    Cost_{machine} (E) = \left(\sum_{i=1}^{n} d_{i}\right) \times M
    \label{eq:cost_machine}
\end{equation}

where $M$ is the constant per-minute price of the infrastructure machines used to run the jobs.

To estimate the total diagnosis cost ($Cost_{diagnosis}$) associated with a category, we hypothesize (from interviews) that the diagnosis process starts after an initial flaky failure and ends with the last rerun (See Figure~\ref{fig:rerun_suite}). Hence, we sum up the time delay required to complete the diagnosis of each initial failure in a category. Then, we multiplied this total time delay by the constant $S$ representing the average cost per minute of diagnosing a flaky failure. For comparative purposes, this constant reflects the hourly rate of a single engineer, even though multiple engineers are often involved in the diagnosis. Accordingly, we estimated the constant based on an average engineer's salary of \$36 per hour, which translates to \$0.6/min. Eq~\ref{eq:cost_diagnosis} shows the formula of the diagnosis cost for a category, described below:

Let $F = \{f_{1},f_{2},...,f_{m}\} \subset E$ ($m < n$) denote the subset of $E$ containing the initial flaky job failures in their respective rerun sequences  $R_{1},R_{2},...,R_{m}$; with $l_{1},l_{2},...,l_{m}$ representing the last executed jobs in each rerun sequence, respectively. 

\begin{equation}
    Cost_{diagnosis} (E) = \left(\sum_{j=1}^{m} Time_{delay}(R_{j})\right)  * S
    \label{eq:cost_diagnosis}
\end{equation}

where $Time_{delay}(Rj) = l_{j}[\mathrm{finished\_at}] - f_{j}[\mathrm{finished\_at}]$ calculated in minutes, and $S$ is the constant average salary of an engineer at the company, estimated per minute.

Finally, we sum the two cost components to calculate the total $Cost$ associated with a category as shown in Eq~\ref{eq:cost_category}.

\begin{equation}
\boxed{
    Cost (E) = Cost_{machine}(E) + Cost_{diagnosis}(E)
}
\label{eq:cost_category}
\end{equation}

To answer our RQ, we rank the failure categories according to the calculated costs and plot the top 20 most costly categories. Our replication package includes additional figures showing the rankings of the remaining categories.

\begin{figure}[tb]
  \begin{center}
      \includegraphics[width=\linewidth]{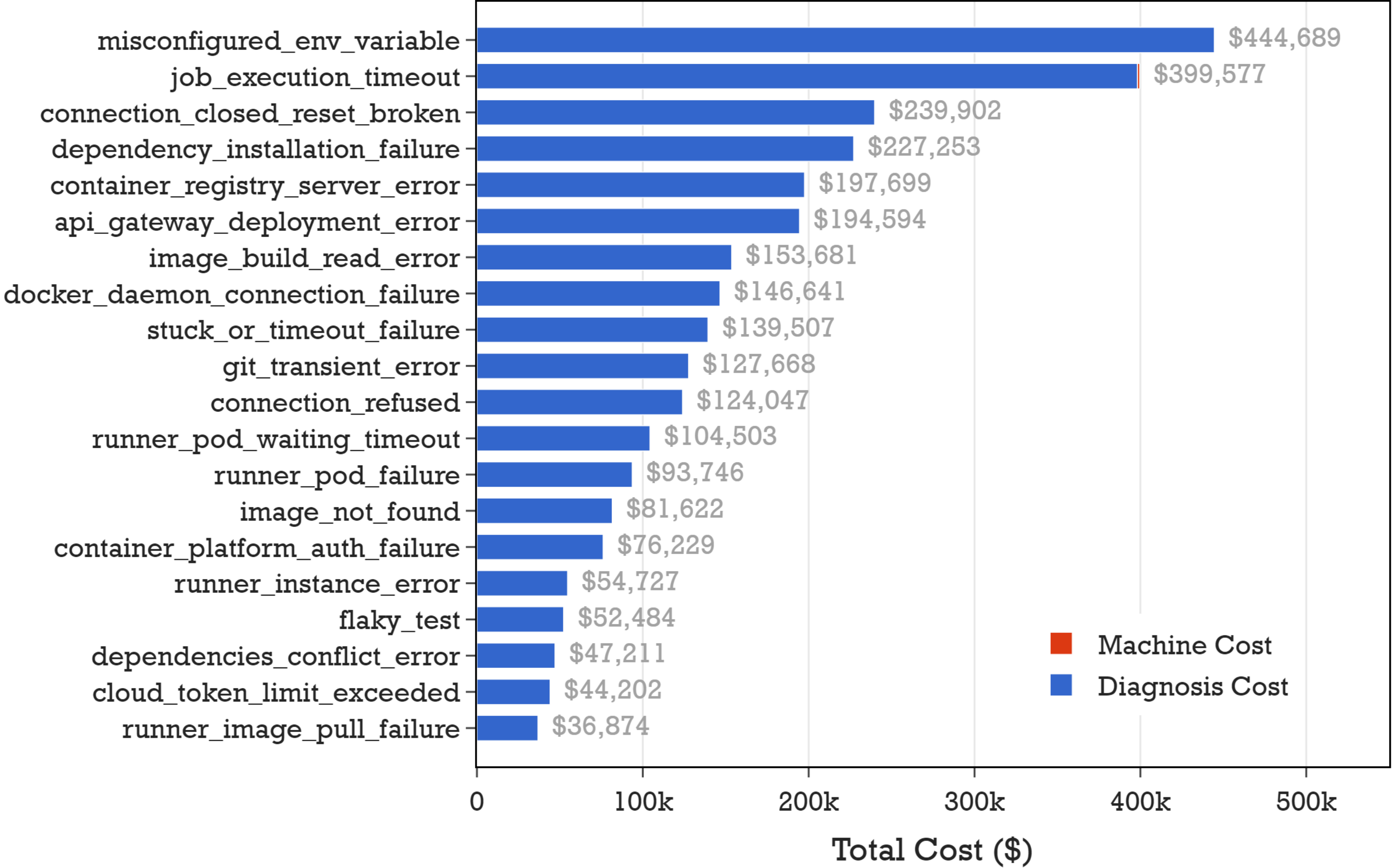}
  \end{center}
\caption{Ranking of the 20 most costly flaky failures categories.}
\label{fig:costly_categories}
\end{figure}

\textbf{Results.} \textbf{The two most costly failure categories are related to environment variable misconfiguration and job execution timeout.} Figure~\ref{fig:costly_categories} displays the 20 failure categories with the highest costs. The most costly category is \textit{misconfigured env variable}, closely followed by \textit{job execution timeout}. These categories are notable as their estimated costs are nearly twice that of the third most costly category. Also, as highlighted in Table~\ref{tab:categories}, they rank first and fourth, respectively, among the most frequent failure categories.

\textbf{Although they occur relatively frequently, the most costly categories are not always the most prevalent.}  The two most costly failure categories, namely \textit{misconfigured env variable} and \textit{job execution timeout}, are followed in the top five most costly respectively by the \textit{connection closed reset broken}, \textit{dependency installation failure}, and \textit{container registry server error} categories. These categories are not part of the top five most frequent ones. However, put together, the five most costly categories account for more than a third (33.69\%) of the labeled flaky failures, with respective proportions of 14.92\%, 6.78\%, 4.52\%, 2.75\%, and 4.72\%. This finding highlights the importance of considering different measures to prioritize failure categories.

Besides, almost the entire cost associated with each category is related to delay times. As it can be observed in Figure~\ref{fig:costly_categories}, the diagnosis cost constitutes the overwhelming majority of the total cost for each of the 20 most costly failure categories, with proportions ranging from 99.63\% for the \textit{flaky test} category to 100\% for the \textit{runner instance error} category. This result underscores the need for automated solutions to address the time-consuming and costly diagnosis (and repair) of flaky failures at TELUS.

\begin{mdframed}
    \textit{\textbf{Answer to RQ2.} The five most costly categories represent over a third (33.69\%) of the flaky job failures and are \textbf{misconfigured env variable}, \textbf{job execution timeout}, \textbf{connection closed reset broken}, \textbf{dependency installation failure}, and \textbf{container registry server error}. The last three are not in the top five most frequent, indicating that cost and frequency are not always directly correlated and should be both considered in prioritization approaches.}
\end{mdframed}

\subsection{\textbf{\rqthree}}

\textbf{Motivation.} The objective of this RQ is to explore the evolution of the failure categories over time. Specifically, we aim to determine whether certain categories have ceased to occur or if new ones have emerged recently. The findings of this RQ will provide valuable insights into the recency and the dynamics of the failure categories, helping to pinpoint the most recent issues affecting the CI system that require attention.

\textbf{Approach.} We identify the distinct creation dates of the jobs in each category. Then, we use a one-dimensional scatter plot to visualize the occurrence dates of the categories across the full 6-year period (indicated in Table~\ref{tab:collected_data}) of the collected data. We intentionally neutralize the daily frequency dimension on the plot since this RQ focuses on temporal trends. To clarify gaps in our visualizations, we also display the evolution of flaky failures with missing logs, which account for a significant proportion of the initial 7,763 flaky failures. For simplicity of the figure, we present only the 20 most frequent categories in this paper. Additional plots for the remaining categories are available in our replication package. 

\begin{figure}[tb]
  \begin{center}
      \includegraphics[width=\linewidth]{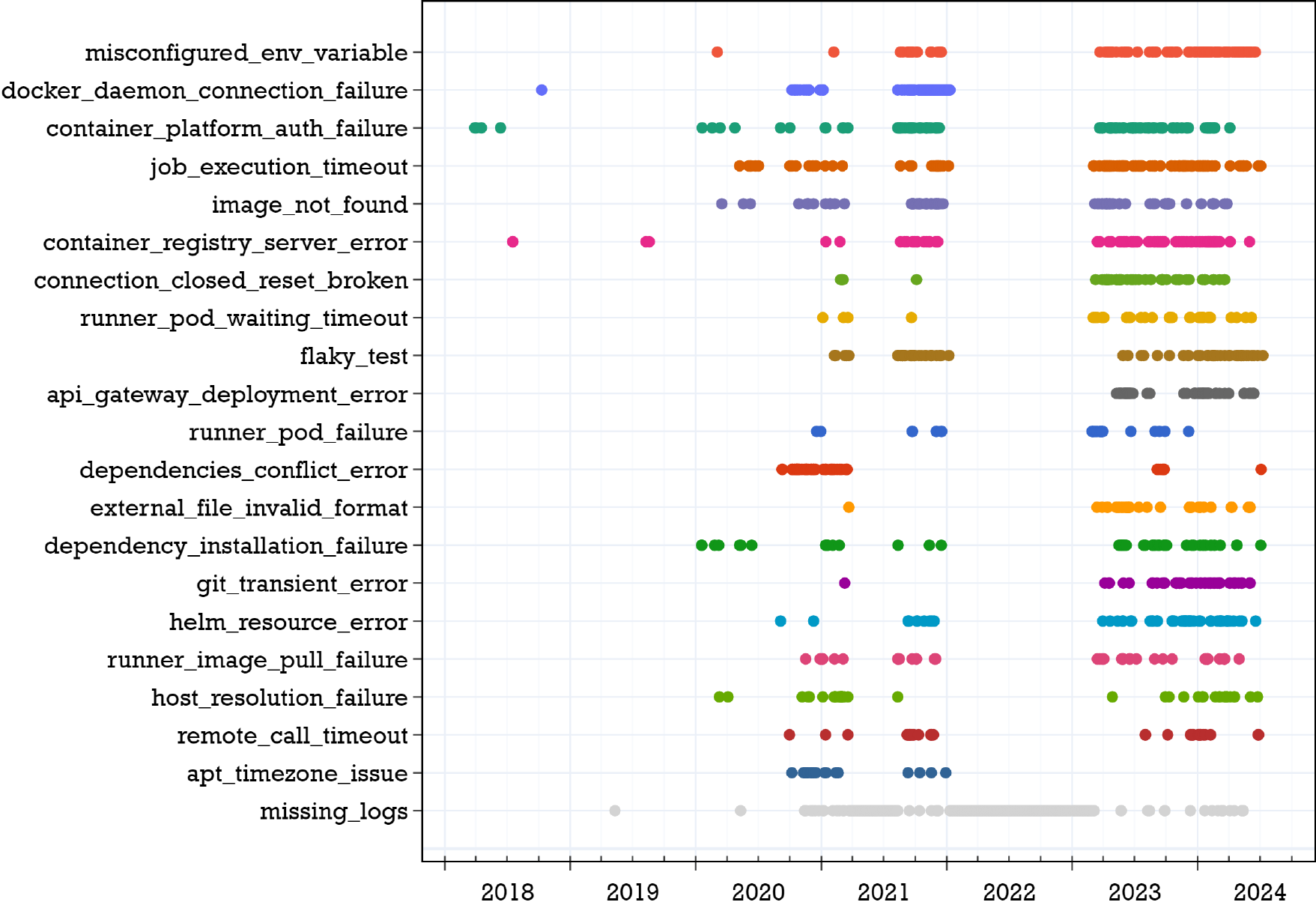}
  \end{center}
\caption{Evolution over time of the 20 most frequent flaky failure categories, arranged from top to bottom. The final scatter plot at the bottom shows the evolution of flaky failures with missing log data.}
\label{fig:categories_evolution} 
\end{figure}

\textbf{Results.} Figure~\ref{fig:categories_evolution} presents the evolution over time of the 20 most frequent flaky failure categories. There are notable gaps in the timeline across all categories during the second quarter of 2021 and again between the beginning of 2022 and the first quarter of 2023, which we refer to as the \textit{big blank period} in the following. These gaps align with the main periods of jobs with missing logs.

\textbf{Some failure categories, including the second most frequent, have not been observed for at least a year.} As shown in  Figure~\ref{fig:categories_evolution}, the categories \textit{docker daemon connection failure} and \textit{apt timezone issue} do not present any data point after the big blank period. Although \textit{docker daemon connection failure} is the second most frequent failure category, all instances of flaky failures identified in this category occurred before February 2022, more than two years before the present study. In addition, while spotted recently, the \textit{dependencies conflict error} category was more regularly observed before the second quarter of 2021. 
This result demonstrates that recency plays a critical role in prioritization, and that frequency alone is insufficient to justify treating a failure category as a priority.

\textbf{Certain failure categories have begun appearing consistently only recently.} Figure~\ref{fig:categories_evolution} illustrates the example of the \textit{api gateway deployment error} category, which is identified only after the big blank period. It also shows that several of the 20 most frequent categories, notably \textit{misconfigured env variable}, \textit{flaky test}, \textit{git transient error}, and \textit{helm resource error} have more consistent apparitions from late 2023 to the second quarter of 2024. These findings suggest that the changes made to the CI/CD system in early 2022 have resolved specific issues while giving rise to new challenges.

\textbf{While some categories have occurred recently, they are not recurrent over the recent period.} As depicted in Figure~\ref{fig:categories_evolution}, the last occurrence dates for the categories \textit{dependencies conflict error} and \textit{remote call timeout} are around mid-2024. However, the dates of occurrence preceding these last appearances are respectively about nine months earlier, at the end of the 3rd quarter of 2023, and five months earlier, in the middle of the first quarter of 2024. This result suggests that using recency based solely on the last occurrence is an ineffective measure for prioritizing a category, whereas recent recurrence provides a more meaningful indicator.

\begin{mdframed}
    \textit{\textbf{Answer to RQ3.} Several failure categories, such as the second most frequent,\textbf{docker daemon connection failure}, and \textbf{apt timezone issue}, have not occurred for over a year. In contrast, other frequent categories, like \textbf{container registry server error} and \textbf{api gateway deployment error}, have regularly emerged only recently after major process changes, showing the importance of considering recent recurrence in prioritization approaches.}
\end{mdframed}

\subsection{\textbf{\rqfour}}

\textbf{Motivation.} The goal of this RQ is to prioritize flaky failure categories based on their studied characteristics: recency, frequency, and cost. The findings will guide the development of an ML model for automated diagnosis and help our industry partner focus on priority failure categories requiring specific repair solutions. Additionally, our prioritization framework can serve as a useful approach for researchers addressing a large variety of failure types in various contexts.

\textbf{Approach.} We build an RFM model \cite{funatsu_data_2011} that takes the R-F-M values of all failure categories as input and applies the K-means \cite{noauthor_kmeans_nodate} clustering algorithm to group them into different clusters. RFM model is a widely used method for behavioral segmentation (clustering) \cite{alves_gomes_review_2023}. The resulting clusters are analyzed to determine which ones should be prioritized. For this purpose, we create a dataset comprising for each of the 46 failure categories, the R-F-M values defined as follows:

\begin{itemize}
    \item Recency (R). We calculate the average number of days that have elapsed since the last three occurrences of the failure category up to the date of the current study. Although traditional definitions of recency only use the last occurrence, we consider the last three in this study to minimize the relative importance of categories that occurred without consistent recurrence in the recent period. For example, we calculated a recency value of 16 and 195 days for the categories \textit{flaky test} and \textit{dependencies conflict error}, respectively, while their last occurrences date back only to 2 and 7 days (as shown in Figure~\ref{fig:categories_evolution}). 

    \item Frequency (F). We use the frequency value of the failure category as reported in Table~\ref{tab:categories}. A higher frequency value indicates that the failure category has been encountered more often by developers.

    \item Monetary (M). We use the monetary cost associated with the failure category. A higher monetary value indicates (based on findings in RQ2) that developers spend a significant amount of time diagnosing this failure category.
\end{itemize}

\begin{table}
\begin{center}
\caption{Flaky Failure Categories with Outliers R-F-M values}
\label{tab:rfm_outliers}
\begin{scriptsize}
\begin{tabular}{lrrrr}
\hline
\textbf{Category}                   & \textbf{R (Days)} & \textbf{F (\#)} & \textbf{M (\$)} & \textbf{Pattern} \\ \hline
misconfigured\_env\_variable        & 27         & 673        & 444,689  & R\textuparrow F\textuparrow M\textuparrow    \\ \hline
job\_execution\_timeout             & 8          & 306        & 399,577  & R\textuparrow F\textuparrow M\textuparrow    \\ \hline
docker\_daemon\_connection\_failure & 915        & 325        & 146,641  & R\textdownarrow F\textuparrow M\textuparrow    \\ \hline    
testing\_device\_oom\_error         & 1,265      & 6          & 6,377    & R\textdownarrow F\textdownarrow M\textdownarrow      \\ \hline
image\_build\_permission\_denied    & 1,415      & 8          & 702      & R\textdownarrow F\textdownarrow M\textdownarrow     \\ \hline 
\end{tabular}
\end{scriptsize}
\end{center}
\end{table}

\begin{table*}[]
\caption{K-Means Clustering of Flaky Failure Categories based on RFM Scores.}
\label{tab:clustering_results}
\centering
\begin{tabular}{lllllllllll}
\hline
\multirow{2}{*}{\textbf{Cluster}} & \multirow{2}{*}{\textbf{Size}} & \multirow{2}{*}{\textbf{Std}} & \multicolumn{2}{l}{\textbf{Recency (Avg.)}}    & \multicolumn{2}{l}{\textbf{Frequency (Avg.)}} & \multicolumn{2}{l}{\textbf{Monetary (Avg.)}} & \multirow{2}{*}{\textbf{Pattern}} & \multirow{2}{*}{\textbf{Description}} \\ \cline{4-9}
                                  &             &                             & \multicolumn{1}{l}{\textbf{Days}} & \textbf{R} & \multicolumn{1}{l}{\textbf{\#}}  & \textbf{F} & \multicolumn{1}{l}{\textbf{\$}} & \textbf{M} &                                   &                                       \\ \hline
C1                                & 6                              & 0.48       & \multicolumn{1}{l}{35.83}         & 4.83       & \multicolumn{1}{l}{164.83}       & 4.67       & \multicolumn{1}{l}{150,700}  & 4.67       & R\textuparrow F\textuparrow M\textuparrow                            & High Priority                         \\ \hline
C2                                & 6                              & 0.45       & \multicolumn{1}{l}{55.83}         & 4.50       & \multicolumn{1}{l}{88.17}        & 3.83       & \multicolumn{1}{l}{29,341}   & 3.17       & R\textuparrow F\textuparrow M\textuparrow                            & Medium Priority                              \\ \hline
C3                                & 5                              & 0.55       & \multicolumn{1}{l}{119.40}        & 3.60       & \multicolumn{1}{l}{47.20}        & 2.60       & \multicolumn{1}{l}{10,970}   & 2.40       & R\textuparrow F\textdownarrow M\textdownarrow                            & Low Priority                            \\ \hline
C4                                & 4                              & 0.51       & \multicolumn{1}{l}{120.50}        & 3.75       & \multicolumn{1}{l}{16.00}        & 1.50       & \multicolumn{1}{l}{743}      & 1.00       & R\textuparrow F\textdownarrow M\textdownarrow                            & Emerging Issues                            \\ \hline
C5                                & 5                              & 0.45       & \multicolumn{1}{l}{162.20}        & 2.80       & \multicolumn{1}{l}{204.40}       & 4.80       & \multicolumn{1}{l}{107,742}  & 4.20       & R\textdownarrow F\textuparrow M\textuparrow                            & Idle             \\ \hline
C6                                & 4                              & 0.44       & \multicolumn{1}{l}{229.75}        & 2.00       & \multicolumn{1}{l}{26.75}        & 2.00       & \multicolumn{1}{l}{117,991}  & 4.75       & R\textdownarrow F\textdownarrow M\textuparrow & Relic                                 \\ \hline

C7                                & 4                              & 0.61       & \multicolumn{1}{l}{471.25}        & 1.75       & \multicolumn{1}{l}{45.00}        & 2.75       & \multicolumn{1}{l}{7,404}    & 2.00       & R\textdownarrow F\textdownarrow M\textdownarrow                            & Negligible                         \\ \hline
C8                                & 7                              & 0.30      & \multicolumn{1}{l}{626.86}        & 1.00       & \multicolumn{1}{l}{9.43}         & 1.14       & \multicolumn{1}{l}{3,736}    & 1.43       & R\textdownarrow F\textdownarrow M\textdownarrow                            & Irrelevant                                \\ \hline
\textbf{Total}                    & 41          &                            & \multicolumn{1}{l}{}              & 3.03       & \multicolumn{1}{l}{}             & 2.91       & \multicolumn{1}{l}{}            & 2.95       &                                   &                                       \\ \hline
\end{tabular}
\end{table*}

 Since K-means is sensitive to outliers, we identify the outliers in our RFM dataset and treat them separately. We use IsolationForest (IF) \cite{noauthor_isolationforest_nodate}, an outlier detection model that leverages a random recursive partitioning technique, resulting in stable outcomes across different initializations. We train the IF model using 500 estimators and a contamination ratio of 10\% 
 to predict the outliers. We repeated the model training and prediction 100 times to ensure the accurate detection of the outliers. Each iteration consistently identified the same five outlier categories reported in Table~\ref{tab:rfm_outliers}. These outliers are removed from the RFM dataset and discussed in the results.

We perform RFM scoring using the quintile method for data normalization. We divide the 41 categories into five equal bins according to the quintiles of the R(F-M) measure. For F(M), a score of 5 is assigned to the top quintile and 4, 3, 2, and 1 to the others. For the R score, 5 indicates a recency value in the lowest quintile. As a result, an RFM score of 555 is attributed to a very recent, frequent, and costly category, while a score of 111 indicates an ancient, rare, and cheap category.

We fit a K-Means model to group categories with similar RFM scores. The model uses \textit{k-means++} initialization to optimize centroid selection. In line with previous work \cite{funatsu_data_2011}, we set $k = 8$ to group the categories into eight distinct clusters for analysis. In fact, each cluster is analyzed by comparing its average R, F, and M values to the overall averages, with an upward $\uparrow$ or a downward $\downarrow$  arrow assigned based on whether the average value of the cluster surpasses or falls below the corresponding overall average (resulting in 8 possible combinations). For instance, the pattern R\textdownarrow F\textuparrow M\textuparrow~ means that the cluster has an average recency below the overall average, while the average frequency and monetary values are above the respective total averages. 

We evaluate the quality of our resulting clusters using the Sum of Square Errors (SSE) \cite{noauthor_error_nodate} and the standard deviation. The K-means algorithm was run 500 times, and the model with the lowest SSE value was selected. The clustering results are then reported to answer this RQ.

\begin{table}
\begin{center}
\caption{Priority Flaky Failure Categories}
\label{tab:priority_categoires}
\begin{scriptsize}
\begin{tabular}{lrrrr}
\hline
\textbf{Category}                   & \textbf{R (Days)} & \textbf{F (\#)} & \textbf{M (\$)} & \textbf{Priority} \\ \hline
misconfigured\_env\_variable        & 27         & 673        & 444,689  & Top    \\ \hline
job\_execution\_timeout             & 8          & 306        & 399,577  & Top   \\ \hline
dependency\_installation\_failure             & 54          & 124        & 227,253  & High    \\ \hline
runner\_pod\_waiting\_timeout             & 35          & 199        & 104,503 & High    \\ \hline
api\_gateway\_deployment\_error            & 29          & 161        & 194,594  & High    \\ \hline
container\_registry\_server\_error             & 41          & 213        & 197,699  & High    \\ \hline
git\_transient\_error             & 40          & 113        & 127,668  & High    \\ \hline
flaky\_test             & 16          & 179        & 52,484  & High    \\ \hline
external\_file\_invalid\_format             & 41          & 125        & 29,672  & Medium    \\ \hline
host\_resolution\_failure             & 24          & 91        & 23,709  & Medium    \\ \hline
runner\_image\_pull\_failure             & 99          & 99        & 36,874  & Medium    \\ \hline
cloud\_token\_limit\_exceeded            & 58          & 39        & 44,202  & Medium    \\ \hline
remote\_call\_timeout             & 61          & 76        & 25,371  & Medium    \\ \hline
helm\_resource\_error             & 52          & 99        & 16,217  & Medium    \\ \hline
\end{tabular}
\end{scriptsize}
\end{center}
\end{table}

\textbf{Results}. Table~\ref{tab:clustering_results} presents the clustering results, describing the eight clusters identified. For each cluster, it shows the size and the average standard deviation (Std) across RFM scores, followed by the average RFM measures and scores. The overall averages of RFM scores are shown in the final row. The table also includes RFM patterns and descriptions of the clusters, ranked from \textit{high priority} to \textit{irrelevant} based on the patterns and average measures. The eight clusters are well-balanced, with sizes ranging from 4 (for clusters C4, C6, C7) to 7 (for C8). Table~\ref{tab:clustering_results} also shows that the clusters are statistically distinct, with average standard deviations ranging from 0.61 being the highest to 0.30 being the lowest.

\textbf{We identified 14 priority failure categories ($\approx$ 30\%) for automated diagnosis and repair solutions}, summarized in Table~\ref{tab:priority_categoires}. Indeed, we prioritize failure categories with high R, F, and M scores and, therefore, associated with the pattern R\textuparrow F\textuparrow M\textuparrow. Hence, we first consider the outliers associated with this pattern. As shown in Table~\ref{tab:rfm_outliers}, \textit{misconfigured env variable} and \textit{job execution timeout} are the top two categories to be prioritized, with the lowest recency measure and highest frequency and monetary measures. These top priority categories are followed by those in clusters C1 and C2, totaling 12 additional categories as shown in Table~\ref{tab:clustering_results}. C1 is ranked as \textit{high priority} due to its average RFM scores being close to 5, which are higher than those of C2. Although C2 follows a similar RFM pattern, it is ranked as \textit{medium priority} due to its comparatively lower average values. The 14 priority failure categories account for 30.43\% of the 46 identified categories.

\textbf{Certain failure categories ($\approx$ 30\%) require close monitoring.} The clustering results show that C3 (R\textuparrow F\textdownarrow M\textdownarrow) comprises less recent categories with low frequency and cost, which we consider to be of \textit{low priority}. Categories in C4 (R\textuparrow F\textdownarrow M\textdownarrow) are even more recent on average with lower frequency and cost, which indicates this cluster may include \textit{emerging issues} (e.g., \textit{runner pod not found}). The categories in these clusters need to be continuously monitored as they reflect current issues that can potentially affect the system more severely over time. In addition, \textit{idle} failure categories in C5 (R\textdownarrow F\textuparrow M\textuparrow) also deserve special attention as they represent highly frequent and costly categories that have not surfaced for a little while. In particular, these categories include recently resolved ones (e.g., \textit{runner pod failure}), for which monitoring is essential to ensure they are fully resolved and do not resurface. C3, C4, and C5 total 14 other categories.

\textbf{The remaining failure categories ($\approx$ 39\%) can be safely ignored}. C6 (R\textdownarrow F\textdownarrow M\textuparrow) contains \textit{relic} categories that were highly costly but are now bygone. The outlier \textit{docker daemon connection failure} (R\textdownarrow F\textuparrow M\textuparrow) also falls into that kind of failure category, which can be analyzed to gain insights into major past issues within the CI system. Clusters C7 and C8 (R\textdownarrow F\textdownarrow M\textdownarrow) group together failure categories described respectively as \textit{negligible} and \textit{irrelevant}. Along with the outliers of the same RFM pattern, i.e. \textit{testing device oom error} and \textit{image build permission denied}, these categories reflect outdated issues that occurred infrequently with minimal costs.

\begin{mdframed}
    \textit{\textbf{Answer to RQ4.} We identified 14 priority failure categories for future automated diagnosis and repair research. The top priority categories, \textbf{misconfigured environment variable} and \textbf{job execution timeout}, stand out as the most prevalent, costly, and recent at TELUS. Also, 14 other categories including idle and emerging issues, have been identified for close monitoring.}
\end{mdframed}

\section{Threats to Validity}

\textbf{Internal Validity.} Internal threats to validity are related to the representativeness of the data collected. To mitigate these threats, we identified key projects where developers are most active and collected flaky failure data from the entire history of these projects. However, approximately 30\% of the flaky failure logs were unavailable. Although TELUS engineers confirmed that the logs were lost due to a migration of GitLab, analyzing flaky failures executed during this period of significant change would enable us to better understand the evolution of issues and draw more accurate conclusions.  

\textbf{Construct Validity.} The primary threat to construct validity lies in the correctness of the identified flaky failure categories. To mitigate this, we used open coding to label a representative sample of logs manually, collaborated with TELUS engineers to resolve disagreements, and developed a regex-based labeling tool that achieved $\approx$ 82\% precision on an unseen sample. This tool is further refined before its use on the dataset. Ultimately, engineers confirmed that the identified categories align with their practical experiences. The second threat concerns the estimation of diagnosis costs (RQ2). Lead time, from the first failure to the final rerun, is used as a proxy of diagnosis time, though it may include idle periods like waiting times and weekends. While this may not perfectly represent diagnosis time, it is the most objective measure available. We accept this limitation since precise cost estimations are less critical for this study, which focuses on prioritization through comparative analysis.

\textbf{External Validity.} The threats to external validity relate to the generalizability of our findings. To mitigate these threats, we analyzed flaky failures from projects varying in size, age, purpose, and programming languages. In addition, TELUS extensively uses open-source and public tools that are widely used across organizations. However, TELUS has a CI environment prone to specific types of flaky failures due to its organization and the nature of its core software products (i.e., SDNs). Therefore, our results may not apply to very different project settings. We are making open sourcing the labeling tool and scripts used to enable the replication of this study in other contexts.

\section{Related Work}

\subsection{Flaky Builds}

In the related work on flaky builds, Durieux et al. \cite{durieux_empirical_2020} analyzed the impact of restarted builds in open-source projects, finding that they reveal flakiness about 47\% of the time, interrupting development workflows and delaying changes merge for up to 48h. Lampel et al. \cite{lampel_when_2021} focused on detecting flaky job failures at Mozilla, using ML algorithms and telemetry data, with factors such as run times and CPU load being key contributors. These factors enabled Mozilla engineers to understand the root cause of flaky failures. Olewicki et al. \cite{olewicki_towards_2022} leveraged textual similarity in job logs to detect brown (flaky) builds at Ubisoft, achieving human-level performance with their ML model. Moriconi et al. \cite{moriconi_automated_nodate} proposed a Knowledge Graph-based approach to detect flaky build failures at Amadeus, outperforming a log similarity-based baseline model.

\subsection{Flaky Test Categories} Several studies focused on categorizing flaky tests to foster research on diagnosing and fixing flaky test failures. Luo et al. \cite{luo_empirical_2014} manually analyzed hundreds of commits data to categorize flaky tests based on the root causes, resulting in a taxonomy of 11 flaky test categories. Later, Eck et al. \cite{eck_understanding_2019} investigated developers' experiences with flaky tests through surveys and interviews to develop a taxonomy that includes four new categories, highlighting the need for better tools to handle flaky tests. These categories have been used in various approaches and tools to diagnose and fix flaky tests. In particular, %
FlakyCat \cite{akli_predicting_2022} was designed based on Few-Shot Learning (FSL) models to classify flaky tests into five categories: \textit{async await}, \textit{test order dependency}, \textit{unordered collection}, \textit{concurrency}, and \textit{time}. Recently, FlakeSync \cite{rahman_flakesync_2024} was introduced to automate the repair of flaky tests in the \textit{async await} category. Similarly, FlakyFix \cite{fatima_flakyfix_2024} applied Few-Shot Learning (FSL) to predict flaky test fix categories, assisting developers in streamlining the repair process.

\section{Conclusion and Future Work}

One of the major challenges affecting CI is flaky job failures. This paper presents an approach to identify and prioritize flaky failure categories for automated diagnosis and resolution. We developed a labeling tool and established a taxonomy of 46 flaky failure categories identified in the TELUS context. We then analyzed the Recency (R), Frequency (F), and estimated Monetary cost (M) associated with each of these categories. This analysis provided valuable insights into how the different categories affect the CI system at TELUS over time. 
It also outlined that R-F-M measures are not correlated and must all be considered when prioritizing categories. Finally, we built an RFM clustering model and identified 14 priority categories for future work on automated diagnosis and repair and 14 others for monitoring as future potential threats.

\section*{Acknowledgment}

We acknowledge the support of the Natural Sciences and Engineering Research Council of Canada (NSERC). TELUS and Mitacs also supported this work through the Mitacs Accelerate program.

\bibliographystyle{IEEEtran}
\bibliography{references}

\end{document}